\documentclass[prl,twocolumn]{revtex4}
\usepackage{graphicx}

\begin{document}
\title{Mechanism of deflagration-to-detonation transition in gas}
\author{Paul Clavin\\
\small Aix Marseille Universit\'e, CNRS, Centrale Marseille,   IRPHE, UMR7342,
49 rue Joliot Curie, BP 146, 13384
Marseille Cedex 13, France\\
}
%%%%%%%%%%%%%%%%%%%%
\begin{abstract}
The deflagration-to-detonation transition on the tip of an elongated flame in a tube is shown to be related to a dynamical saddle-node bifurcation of the inner flame structure leading to a runaway of the pressure in finite time. The comparison with the experiments shows a good agreement.
\end{abstract}
\maketitle
%%%%%%%%%%%%%%%%%%%%%%%%%%
\subsection{Introduction and orders of magnitude}
The deflagration-to-detonation transition (DDT) is a quasi-instantaneous transition (few microseconds) between two opposite regimes of combustion wave. The phenomenon was observed long ago \cite{shcheltrosh}. However, after more than a century of experimental works and decades of numerical studies reported in an extensive literature, DDT is not yet understood 
\cite{lee2008}\cite{Clavin2016}. 
This abrupt transition is explained here by the one-dimensional dynamics of the reactive flow of a self-accelerating flame. Coupling reaction-diffusion and compressibility is a challenging problem. The physical mechanism of the DDT is presented here in a synthetic way for physics-oriented readers, skipping the technical details of the analytical methods that will be published elsewhere \cite{Clavin2023}. 
Gaseous detonations are supersonic combustion waves involving a  pressure rise $\Delta p/p$ ranging from 20 to 50 and propagating with a velocity between 2\,000 m/s  and 3\,500 m/s under normal conditions. Laminar flames are quasi-isobaric reaction-diffusion waves  characterized by a markedly subsonic velocity and a negligible pressure drop $\Delta p/p\ll 1$. In gas, the laminar flame velocity relative to the unburned gas $U_L$ ranges from few tens cm/s in the least energetic mixtures to 9 m/s in the most energetic ones. 
 The rate of heat release is governed by inelastic collisions of molecules associated with an activation energy $\mathcal E$ larger than the thermal agitation $k_BT$. The flame temperature $T_b$ and the reduced activation energy $\beta\equiv \mathcal E/k_B T_b$ are in the ranges 1800 to 3500 K and 2 to 8 respectively; the larger $T_b$, the smaller $\beta$. According to the kinetic theory of gas, the reaction rate $1/\tau_r$ is expressed in terms of the collision frequency $1/\tau_{coll}$ in the form of an Arrhenius law  $\tau_{coll}/\tau_r\propto  \beta\text e^{-\beta}$. However, due to a complex chemical kinetics network \cite{SaFaw}, the exothermic reaction rate decreases drastically  below a crossover temperature $T_c\in [850\,-1200\,K]$ and combustion cannot develop for $T<T_c$. The reaction rate $1/\tau_r$ at $T_b$ is typically  $1/\tau_{rb}\approx 10^{6}$\,$s^{-1}$.  
 Detonations consist of a strong inert shock of few mean free paths thick ($\approx a \tau_{coll}$, $a$ is the sound speed) followed by a macroscopic zone of reaction of thickness $\approx a \tau_r$  through which the diffusive transports are negligible  $\tau_r\gg \tau_{coll}$.  Expressed in terms of the molecular diffusivity $D\approx a^2 \tau_{coll}$, according to the  Zeldovich-Frank-Kamenetstskii (ZFK) analysis \cite{ZFK} in the limit $\beta\gg 1$,  the laminar flame velocity relative to the burned gas is $U_b\propto \beta^{-3/2}\sqrt{D_b/\tau_{rb}}$ where the subscript b refers to the burned gas, see \cite{Clavin2016} for a didactic presentation. The flame thickness and the transit time of a fluid particle across the flame are respectively $d=D_b/U_b\approx 2-4\times10^{-1}$mm and $t_{b}=d/U_b\propto \beta^{3}\tau_{rb} \approx 10^{-4}$s under normal conditions. 
The flame Mach number $\varepsilon\equiv U_b/a_b \propto \text e^{-\beta/2}/\beta$ is about $0.03-0.1$ in very energetic mixtures (stoichiometric $H_2/O_2$ or $C_2H_4/O_2$ mixtures) and $10^{-3}$ in hydrocarbon-air mixtures and  
$\Delta p/p\approx \varepsilon ^{2}$. 
%%%%
\subsection{Experimental data and objective of the analysis} 
%%%%
We limit our attention to 
DDT in long smooth-walled tubes filled with  stoichiometric $H_2/O_2$ or $C_2H_4/O_2$ mixtures.
Just before transition in micro-scale tubes (radius $\approx$ few mms) the front of the laminar flame is elongated and the flow is laminar \cite{Wu10}\cite{Bikov22}, see {fig.1}. 
\begin{figure}[h]
\includegraphics[width=8cm]{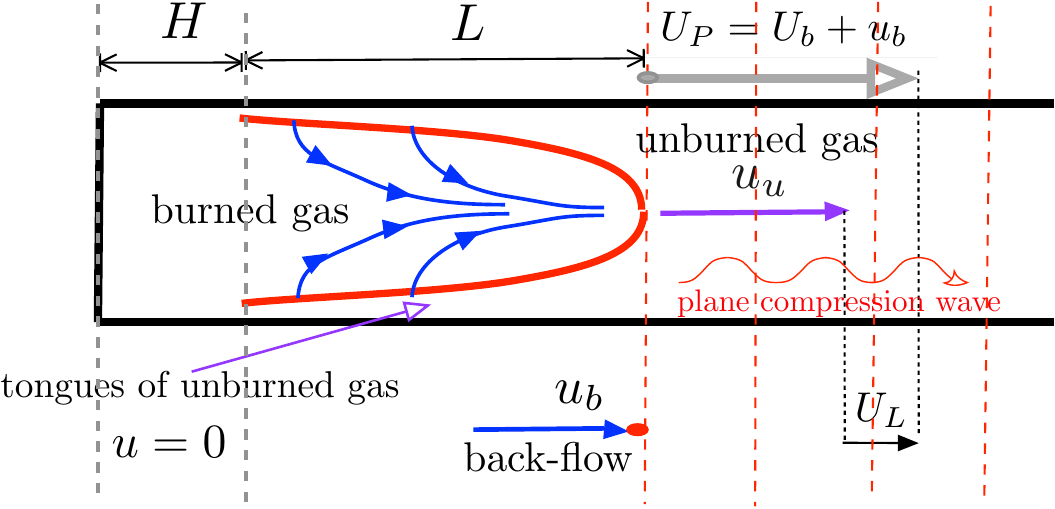}
\caption{Sketch of an elongated flame in a smooth-walled tube }
\label{fig.1}
\end{figure}
The DDT is more difficult to analyze in larger tubes because the flame front can be turbulent  \cite{Opp66}, the problem becoming stochastic in nature. Conclusions can nevertheless be drawn 
since 
the salient features of the abrupt transition  in moderately large tubes (50\,x\,50\,mm$^2$) are not so different, 
especially   
as the flow stays laminar ahead of the wrinkled flame front \cite{Lieb10}\cite{Kuz10}. The salient
 features can be summarized as follows: spontaneous onset of detonation occurs  locally outside the boundary layers in a small explosion center on a laminar flamelet near the leading edge of the flame brush. In some turbulent flames \cite{Opp66} the explosion center is in the boundary layer where the DDT could be due to a Zeldovich mechanism \cite{Zeldo1980} reinforced by the compressible waves \cite{Clavin2016}. This case is not considered here where the attention is focused on an inherent DDT mechanism of laminar flames accelerated by a self-generated convection flow. In a tube of diameter between 5 and 50 mm in which the reactive mixture is ignited at the center of the closed end, the surface area of the laminar flame first increases exponentially, goes through a transient tulip-shaped form \cite{Searby96} and then increases slowly before the abrupt transition. Correlatively, a longitudinal convection flow is generated by the  increase of surface-area  of the flame front, putting the overall flame in motion at velocity $U_P>U_b$. Because of the laminar nature of the flamelets, the longitudinal flow of unburned gas relative to a flamelet   orthogonal to the tube axis is equal to the laminar flame velocity $U_L\ll a$.  
 The  detonation 
onset occurs spontaneously when appropriate local conditions are attained in the unburned gas adjacent to the leading edge of the self-accelerating flame \cite{lee2008}\,\cite{Lieb10}.  This occurs when the overall speed of the flame front reaches a critical value $U_P^*$ comparable to the local sound speed \cite{Bikov22}\cite{Lieb10}\cite{Kuz10}.  
The acceleration-induced compression waves warm up the unburned gas by compressional heating but, as noticed a long time ago, the temperature stays well below the crossover temperature $T_c$ so that the reactive mixture just ahead of the tip of the elongated flame remains chemically frozen leaving unexplained the DDT. 
We will not discuss the origin of the growth of surface area $\Sigma(t)$ responsible for accelerating the flame front. The flame elongation in a tube with stick walls is governed by the quasi-isobaric expansion of the gas across the flame front  while the abrupt transition under consideration is a consequence of the compression waves in the external flow. In the following, a small growth rate of the total surface area $\epsilon\equiv (t_b/\Sigma )\text d\Sigma /\text dt\ll 1$ is a given parameter (not to be confused with $\varepsilon$). The analysis of the accelerating flame on the tip of the elongated front sketched in {fig.1} is performed  in the limit of small flame Mach number $\varepsilon\equiv U_b/a_b\ll \epsilon\ll1$ using the ZFK flame model. 
 The objective is twofold. Firstly, identify the critical condition from  which the slightest increase in elongation, no matter how small, produces the compressibility-induced DDT onset. This part is an extension of a recent work in which the self-accelerating flame was treated as a planar discontinuity \cite{Clavin2022}, as in the pioneering analysis identifying a turning points of the self-similar solutions of wrinkled flames \cite{Joulin89}. 
 The second objective is to show that the DDT corresponds to a finite-time singularity of the inner structure of an accelerating quasi-planar flame orthogonal to the direction of propagation and put in motion by the convective flow. 
The analysis takes advantage of three simplifications: firstly,  a local  quasi-planar geometry on the tip (back-flow model  \cite{Clavin2021}), secondly, a negligible pressure drop across the flame $\Delta p/p\approx \varepsilon^2 $ (smaller than the pressure variation of order $\varepsilon $ across the compression waves in the external flow), and finally a scale separation, the laminar flame thickness $d$ being much smaller than both the tube radius, $d/r<10^{-1}$ and the length $l$ involved in the compression waves, $d/l < \varepsilon$.

\subsection{Quasi-steady self-similar solution}     

To begin, consider the acceleration-induced compression waves in the unburned gas,  neglecting unsteadiness elsewhere. Due to a quasi-isobaric gas expansion, the longitudinal flow increases across the flamelet, from $u_b$ on the burned gas side to $u_u$ on the unburned gas side, $u_u>u_b$. If the inner structure is in steady state, mass conservation yields $u_b+U_b=u_u+U_L=U_P$ where $U_P$ is the nearly sonic speed of the convected flame front while $U_b$ and $U_L$ are the laminar flame velocities (markedly subsonic) relative respectively to the burned and unburned gas, $\rho_b U_b=\rho_uU_L$ (mass conservation) and $\rho_b/\rho_u=T_u/T_b\in{[0.13-0.08]}$ (quasi-isobaric condition), subscripts u and b denoting respectively the unburned gas ahead of the flame and the burned gas behind the flame. If the inner flame structure is the same everywhere on the elongated front (this is not an essential assumption as discussed few lines below (\ref{elong-})), the mass of fresh gas burned by unit-time yields $\rho_uU_L\Sigma= \rho_bU_b\Sigma$.  Assuming that the flow is at rest  between the closed end of the tube and the straight-section delimiting the  burned gas inclosed in the finger flame,  the mass rate of burned gas is $\rho_b(\text d H/\text dt)\Phi$, $\Phi$ and $H$  denoting respectively the cross section of the tube and the tube length behind the finger flame ($\text d H/\text d t\approx U_P$), see {fig.1}. This  yields the classical result $U_P\approx (\Sigma/\Phi)U_b$. The relations  $u_b=U_P-U_b$ and $u_u=u_b+U_b-U_L$ then lead to the flow velocities $u_b$ and $u_u$ on the flamelet in terms of $U_b$ and the flame elongation $S\equiv \Sigma/\Phi-1$ ($S\approx 2L/R$  for an elongated flame of length $L$ in a cylindrical geometry),
\begin{eqnarray}
U_P=(S+1)U_b, \quad u_b=SU_b, \quad u_u=(S+1-\theta_0)U_b
\label{elong-}
\end{eqnarray}
where $1-\theta_0$ is the reduced heat release, $\theta_0 \equiv T_{u0}/T_{b0}<1$, the subscript $0$ denoting the initial condition. The greater the flame surface area $\Sigma$, the larger the longitudinal flows $u_b$ and $u_u$ on the flamelet.
When approaching the abrupt transition, Eqs. (\ref{elong-}) are meaningful  independently of the precise definition of $S$ (or the length $L$ of the elongated flame). The crucial points are $u_b\propto U_b$, $u_u\propto U_b$ and a rate of change of $U_b$ which diverges  at the transition, no matter how small $\text d S/\text d t$, see (\ref{singu1}) below. Therefore, the rate of change of $S$ does not matter. 
Consider an elongated flame accelerating from an initial state (subscript $0$) constituted by a self-similar solution with the lead shock at infinity; $t\le0:$ uniform flow of unburned gas, $u_u=u_{u0}, \, p=p_{u0}$.  
The flame acting as a semi-transparent piston,  the downstream-running compression waves ahead of the flame are isentropic 
 as long as the variation of $U_L$ is not larger than its initial value 
 because, the acceleration-induced shocks being weak for $\varepsilon \ll 1$, 
$\delta u_u/a_{u0}=O(\varepsilon)$, $\delta p_u=\rho_{u0}a_{u0}\delta u_{u}$, $\delta p_u/p_{u0}=\gamma\delta u_u/a_{u0}\propto \varepsilon$,    
the entropy jump of  $\varepsilon^2$-order is negligible.   
 Denoting $x=X_p(t)$ the position of the flame  $\text dX_P/\text d t=U_P(t)$ and $u_u(t)$ the flow just ahead of the flame, the compressible flow of  unburned gas is, $\varepsilon \ll 1$:
 \begin{eqnarray}
0< x-X_P(t)< a_{u0} t&:& u=u_u\big (t-\big [x-X_P(t) \big]/a_{u0} \big),
\nonumber \\  
a_{u0} t < x-X_P(t)&:&  u=u_{u0}.
\label{acous1}
\end{eqnarray}
The instantaneous  pressure  and temperature  of unburned gas on the flame $\delta p_u(t)$ and $\delta T_u(t)$ are expressed in terms of $u_u(t)$ by the acoustic relations $\delta p_u/p_{u0}=\gamma\delta u_{u}/a_{u0}$ and $\delta T_u/T_{u0}$ = $(\gamma-1)\delta u_u/a_{u0}$.  Using  $U_{b0}/a_{u0}=\varepsilon/\sqrt{\theta_0}$ where  
$a_{b0}/a_{u0}=1/\sqrt{\theta_0}$, one gets
\begin{eqnarray}
\frac{\delta p_u}{p_{u0}}= \frac{\gamma}{\sqrt \theta_0}\varepsilon\,\frac{\delta u_u(t)}{U_{b0}}, \quad  \frac{\delta T_u}{T_{b0}}=(\gamma-1)\sqrt{\theta_0}\,\varepsilon\,\frac{\delta u_u(t)}{U_{b0}}.
\label{acous2}
\end{eqnarray}
For $\delta u_u(t)/U_{b0}$ not larger than unity, the pressure variation on the flame is larger than the pressure drop across the flame by a factor $1/\varepsilon$. Neglecting $\varepsilon$-terms, the flow velocity in the steady inner structure of the flame is the same as in steady state $u_u-u_b=U_b-U_L\approx  (1-\theta_0) \,U_b$.  
Then, according to (\ref{elong-}) $u_b=SU_b$, $u_{u0}=(S_0+1-\theta_0)U_{b0}$,
\begin{eqnarray}
{\delta u_u}/{U_{b0}}+(S_0+1-\theta_0)=(S+1-\theta_0){U_{b}}/{U_{b0}}.
 \label{ub-uu}
\end{eqnarray}
 Under the steady state approximation of the inner structure of the flame, the flame temperature is simply shifted by the temperature change of the unburned gas just ahead of the flame $\delta T_b=\delta T_u$. An  equation for $\delta T_b/T_{b0}$ is obtained  
 by introducing the last equation (\ref{acous2}) into (\ref{ub-uu}), using the expression ${U_{b}}/{U_{b0}}$ in terms of $\delta T_b/T_{b0}$ given by the flame theory.  The ZFK result $\beta\gg 1$: ${U_{b}}/{U_{b0}}\approx$   $\text e^{(\beta/2)\delta T_b/T_{b0}}$, $\delta T_b/T_{b0}=O(1/\beta)$,  yields  the   equation  previously obtained with a flame treated as a discontinuity \cite{Clavin2022}. 
Introducing the  reduced flame temperature $\vartheta\ge 0$ (and/or the flow just ahead the flame ${u_u}/{U_{b0}}$)  and  the reduced elongation  $\sigma\ge1$ 
\begin{eqnarray}
\vartheta \equiv \frac{\beta}{2}\frac{\delta T_b}{T_{b0}}=b\frac{\delta u_u}{U_{b0}}, \qquad
\sigma \equiv \frac{S+1-\theta_0}{S_0+1-\theta_0},
\label{Tbuu}
\end{eqnarray}
the function $\vartheta$ versus $\sigma$ is given   by the roots of
\begin{eqnarray}
&&\qquad \qquad \qquad  \sigma\,\text e^\vartheta-{\vartheta}/{s_0}-1=0,
\label{yxi}
\\
&& s_0\equiv b(S_0+1-\theta_0)\le1, \quad b\equiv (\varepsilon\beta/2)(\gamma-1)\sqrt\theta_0,\qquad
\end{eqnarray}
where the parameter $s_0$ characterizing the initial elongation $S_0$ ($\sigma_0=1$) is smaller than unity $s_0\le1$ because the parameter $b$ is very small, ranging  from $ 1.2\,10^{-2}$ in a stoichiometric H$_2$-O$_2$ flame to $6\,10^{-4}$ in hydrocarbon-air flames, $b\ll 1$ and  $s_0\approx bS_0 \le1$ of order unity. According to (\ref{yxi}), $\text d \sigma/\text d \vartheta=$ $[(1-s_0)-\vartheta]/s_0$, the maximum of the curve $\sigma$ versus $\vartheta$ (turning point of $\vartheta(\sigma)$) defines a critical flame temperature $\vartheta^*=1-s_0$ and a critical elongation $S^*\ge S_0$, $\sigma^*=\text e^{(s_0 -1)}/s_0\ge 1\,\forall s_0$, above which Eq. (\ref{yxi}) has no root. The temperature of the downstream-running compression waves in the unburned gas  for $\sigma>\sigma^*$ is  no longer compatible with the temperature of the accelerating flame acting as a semi-transparent piston. Such a drastic effect associated to a nonlinear thermal feedback was identified long ago in a different context \cite{Joulin89}.  It can be illustrated simply by the compression waves issued from an impermeable piston  whose velocity is an increasing function of the gas temperature \cite{Clavin2021}. 
The critical elongation $S^*$ of (\ref{yxi}) increases with the initial one $S_0$, $\text d \sigma^*/\text d s_o=(1-s_0)\text e^{(s_0-1)}/s^2_0>0$ ($0<s_0\le 1$).  There is a common upper bound   $S^*_{max}=S_{0\,max}\approx 1/b$ ($s_0=1$: $\text d \sigma^*/\text d s_o=0$, $ \sigma^*=1$) and the critical Mach number $(u_u/a)^*_{max}\approx 1/[(\gamma-1)\beta]$ depends only on the thermal sensitivity of the flame and is close to unity as in the experiments  \cite{Bikov22}\cite{Lieb10}\cite{Kuz10}. 
Eq. (\ref{yxi}) has two solutions for $\sigma <\sigma^*$: 
 $\vartheta_-(\sigma)<\vartheta^*$ and  $\vartheta_+(\sigma)>\vartheta^*$, $\vartheta_+(\sigma^*)=\vartheta_-(\sigma^*)=\vartheta^*$,   $\text d\vartheta_+/\text d \sigma <0$ and $\text d\vartheta_-/\text d \sigma >0$. According to the thermodynamic laws, the temperature increases by compression, so that $\vartheta_-(\sigma)$ is
the physical branch.
The smallest critical elongation $S^*_{min}$ is obtained for $S_0=0$, $s_0=b(1-\theta_0)\ll 1$, $\sigma^*=\text e^{(s_0 -1)}/s_0 \approx 1/(s_0\text e)$, $S^*_{min}\approx  S^*_{max}/\text e\approx (1-\theta_0)/(s_0\text e)$. Such critical elongations are too large (at least by a factor $10^2$) for the DDT of laminar hydrocarbon-air flames ($b\approx 6\,10^{-4}$) to happen in smooth-walled tubes, as it is well known.  For very energetic mixtures  ($b\approx 1.2\,10^{-2}$) one gets $L^*/R\in [15, 40]$,  still larger than in experiments but only by a factor 5. The discrepancy comes from the real chemical kinetics of H$_2$-O$_2$ flames \cite{SaFaw} which cannot be represented in a large range of temperature by an Arrhenius law with a single activation energy. However, with the ZFK result extended to $\beta=2$ in the form ${U_{b}}/{U_{b0}}\approx$  $(T_b/T_{b0})^3$ $\text e^{(\beta/2)\delta T_b/T_{b0}}$,  Eq. (\ref{ub-uu}) yields critical elongations similar to the experimental data $L^*/R\approx 3$ to $10$. 

Expanding Eq.(\ref{yxi}) near the critical point ($\vartheta^*$, $\sigma^*$), $\text d \sigma/\text d \vartheta\vert_{\vartheta=\vartheta^*}=0$, yields a  quadratic form 
\begin{eqnarray}
\frac{(\sigma-\sigma^*)}{\sigma^*}+\frac{(\vartheta-\vartheta^*)^2}{2}=0,\,\, \,\,  \vartheta^*-\vartheta_-=\sqrt 2\sqrt{\frac{\sigma^*-\sigma}{\sigma^*}},
\label{yxigene}
\end{eqnarray}
showing how 
the flame temperature $T_b$ (or the convective flow $u_u$)
behaves when the elongation approaches the critical value $\sigma^*$. The key point is the singular derivative $\text d \vartheta_-/\text d \sigma$ 
when $\sigma \to \sigma^*$ from below. Introducing  the non-dimensional rate of elongation $\epsilon$ 
and denoting $t^*$ the time at which the critical elongation $\sigma^*$ is reached ($t\le t^*$)
\begin{eqnarray}
\epsilon\equiv \frac{t_b}{\sigma^*}\frac{\text d \sigma}{\text d t}\Big \vert_{t=t^*}, \,\,\,\, \text{Eq. }(\ref{yxigene}) \,\, \Rightarrow \,\,t_b\frac{\text d \vartheta_-}{\text d t}= \frac{\sqrt{\epsilon/2}}{\sqrt{(t^*-t)/t_b}}
\label{singu1}
\end{eqnarray}
 the time derivative of the flame temperature increases strongly when approaching $t^*$  ($t^*-t\approx \epsilon \,t_b$) and diverges in finite time $t=t^*$. Smaller  the elongation  rate $\epsilon$, sharper the runaway of $\text d \vartheta_-/\text d t$. 
 \subsection{Finite time singularity and DDT mechanism}
Equation (\ref{singu1}) corresponds to a finite time singularity of the flame acceleration  while the critical velocities $u_b^*$,  $u_u^*$ and $U^*_b$ stay finite. In fluid mechanics a shock wave is spontaneously generated on the piston when the acceleration becomes singular. This 
 suggests the formation of a shock inside the inner structure of the laminar flame at $t=t^*$. The phenomenon is not that simple. As shown in the next paragraph, due to unsteady effects not included in (\ref{singu1}), the DDT mechanism is associated with a run away of the pressure and temperature occurring  at $t$ slightly above
$t^*$.  The pressure being quasi-uniform inside the inner flame structure, the latter is blown off as a whole at finite time, leading to a sharp decrease of the flame thickness followed quasi-instantaneously by the detonation offset. Involving time and length scales of the order of the mean free path and collision time respectively, neither the formation of the lead shock of the detonation nor the transient  overshoot of pressure and/or temperature cannot be accurately described by the macroscopic equations of fluid mechanics. However, once the overdriven detonation is formed, the decay to the Chapman-Jouguet regime can be properly obtained by treating the lead shock as a discontinuity across which the Rankine-Hugoniot jump conditions are applied. Notice that the singularity is not due to the nonlinear wave breaking usually involved in the formation of shock waves but by the finite-time divergence of the flame acceleration in (\ref{singu1}), as said before.
\subsection{Dynamical saddle node bifurcation induced DDT}
 
 The understanding of the abrupt transition requires to perform an unsteady analysis of  the full reactive flow, not only  of the downstream-running compression waves  ahead of the flame described in (\ref{acous1}).
 Two additional unsteady effects have to be investigated, one concerning the flow of burned gas, and the other the inner  structure of the laminar flame. 
 Due to the complexity of the curved flow of burned gas sketched in  {fig.1}, the first unsteadiness can be only roughly evaluated; hopfully a detailed solution is not actually needed. The back-flow $u_b(t)$ on the tip of the elongated flame results from a cumulative effect of burned gas flows issued from the lateral flames that are fed by the fresh gas of the tongues quasi-parallel to the wall,
  $u_b(t)=(2/R)\int_{X_P-L}^{X_P}U_b( t-\Delta t_{bg})\text d \text x $ where $U_b(t)$ is the laminar flame velocity on the tip and $\Delta t_{bg}(\text x)\propto (X_P-\text x)/2a$ 
  is the delay 
for transferring to the tip the transient flow of burned gas of the lateral flame at a distance $X_P-\text x$ from the tip.   
For a  change in laminar flame speed sufficiently slow $(L/a)\text dU_b/\text d t\ll U_b$, $(t_b/U_b)\text dU_b/\text d t=O(\epsilon)$, the instantaneous model (\ref{elong-}) $u_b(t)=S(t)U_b(t)$ is replaced  by a delayed back-flow evaluated by integrating $\Delta t_{bg}(\text x)$
\begin{eqnarray}
u_b(t)\approx S(t)[U_b(t)-\Delta t_{w}\text dU_b/\text d t], \quad \Delta t_{w}\approx L/(2a)
\label{delflamod}
\end{eqnarray}
where $L^*/(2a t_b)<1/\epsilon$ is anticipated, $L^*/d<2/(\varepsilon\epsilon)$. 

The unsteady analysis of the ZFK flame structure is performed in the limit $\varepsilon\ll \epsilon \ll 1$, $b=O(1)$,  retaining unsteady terms of order $\varepsilon \epsilon$ while neglecting terms of order  $\varepsilon ^2$. This leads to an equation similar to (\ref{ub-uu})
\begin{eqnarray}
(u_u-u_b)/U_{b0}= (1-\theta_0)\text e^{\beta(\theta_b-1)/2}[1+O(\varepsilon)]
\label{ub-uu2}
\end{eqnarray}
 but where  the flame temperature $\theta_b(t)\equiv T_b(t)/T_{b0}$ is no longer related to the instantaneous value of the flow in the unburned gas $u_u(t)$ by the first relation in (\ref{Tbuu}). The unsteadiness induced variation of flame temperature $\delta \theta_b(t)\equiv \delta T_b(t)/ T_{b0}=O(\varepsilon \epsilon)$ and of mass flux across the reaction sheet $\delta \text e^{\beta(\theta_b-1)/2}=O(b\epsilon)$ are obtained from the unsteady analysis of the inner flame structure. This  involves technical difficulties but the unsteady effects 
are easily understood: according to the solution of the energy equation perturbed by a transient compressional heating, heat conduction  introduces a time delay $\Delta t_{\theta}$ proportional to the transit time across the flame $\Delta t_{\theta}/ t_b= O(1) $, the compressible transfer, $(d/a_b)/(d/U_b)\approx \varepsilon$ being negligible. Therefore the instantaneous mass flux across the reaction sheet differs from the quasi-steady solution by a term proportional  to the time-derivative of the pressure $(\text d p_u/\text d t)/p_{u0}=\gamma(\text d u_{u}/\text d t)/a_{u0}=O(\varepsilon \epsilon/t_b)$.  This introduces a derivative-term proportional to $(\Delta t_{\theta}/U_{b})\text d  u_u/\text d t=O(\epsilon)$ into the right-hand side  of (\ref{ub-uu2}). Introducing $\delta u_u(t)$ given by  (\ref{acous2}) and the delayed back-flow (\ref{delflamod})  $\delta u_b(t)/(SU_{b})=-(\Delta t_{w}/U_{b})\text d  U_b/\text d t=O(\epsilon)$ into the left-hand side of (\ref{ub-uu2}) yields a dynamical equation for the flame temperature. Near the critical point, one gets the first equation (\ref{yxigene}) completed by 
a time-derivative term $  s_oK \text d \vartheta/\text d t$ on the right-hand side with a  non-dimensional coefficient $K\propto (\Delta t_{w}-\Delta t_{\theta})/t_b$.  
Using $(\sigma-\sigma^*)/\sigma^*=\epsilon (t-t^*)/t_b$, the so-obtained equation characterizes  a dynamical saddle-node bifurcation. Conveniently rescaled, $\vartheta-\vartheta^* \to y$ $\propto (\sigma^*/\epsilon s_0\vert K \vert )^{1/3}(\vartheta-\vartheta^*)$, $(\sigma-\sigma^*)/\sigma^* \to \tau\propto (\sigma^*/\epsilon s_0 \vert K\vert )^{2/3}(\sigma-\sigma^*)/\sigma^*$, this equation  takes a normal 
  form extensively used 
  for  sharp transitions in physics, biophysics and catastrophic events \cite{Pomeau2012}
  \begin{eqnarray}
\tau + y^2=\pm \text dy/\text d \tau
\label{sdbif}
\end{eqnarray}
 where $\pm$ is the sign of the coefficient $K$ and $\tau=0$, $y=0$ is the turning point. 
The trajectories in the phase space show that the branch of negative quasi-steady-solutions $\overline y_-(\tau)=-\sqrt{-\tau}$ is stable with the $+$ sign ($K>0$) and unstable with the $-$ sign ($K<0$). Therefore, the unsteady inner structure of the flame contributing to $K$ by a negative term $-\Delta t_{\theta}/t_b<0$, destabilizes the flame subjected to an instantaneous back-flow.  The branch of physical solutions (\ref{yxigene}) is stabilized if the delay introduced by the flow of burned gas is longer than the transit time across the inner structure  
$\Delta t_{w}>\Delta t_{\theta}=O(t_b)$, $K>0$. 
This is the case if the finger flame is much longer than the laminar flame thickness, $L/d>1/(\text e\,\varepsilon)$ ($L\ge 1$ cm in energetic mixtures). The key point of the saddle-node bifurcation (\ref{sdbif}) is that, starting from the stable branch $\tau=\tau_i<0$: $ y=-\sqrt{-\tau_i}$,  $\vert \tau_i\vert >1$, the solution is blown off  at $\tau=\tau_c\approx2.338$, $y(\tau)\approx 1/(\tau_c-\tau)$ \cite {Pomeau2012}, in contrast to $\overline y_-(\tau)$. Unsteadiness makes the entire laminar flame structure blown off few time after the turning point. The smaller the elongation rate $\epsilon \ll 1$, the closer to the turning point the pressure runaway (DDT onset). 

\subsection {Limitation of the analysis and perspectives}

Equation (\ref{sdbif}) is  obtained by an expansion around the turning point. Therefore, the asymptotic behavior  $y(\tau)\approx 1/(\tau_c-\tau)$ for $\tau=\tau_c\approx2.338$ is not guaranteed for the exact solution of the macroscopic equations satisfying the boundary conditions ahead of and behind the self-accelerating flame under the effect of the bak-flow. Nevertheless, the onset of a finite time singularity  is not doubtful because terms of higher order  than in (\ref{sdbif})  are expected to reinforce the singularity. This is illustrated by the divergence of the acceleration $\text d \vartheta/\text d \tau\propto 1/\sqrt{\tau^*-\tau}$ for a flame structure in steady state becoming $\text d \vartheta/\text d \tau\propto 1/(\tau_c-\tau)^2\propto (\vartheta_c-\vartheta)^2$ when a weak unsteadiness is taken into account. Moreover, when a second derivative term $\text d^2 y/\text d \tau^2$ becomes larger than the first derivative on the right-hand side of (\ref{sdbif}) the asymptotic solution of such a first Painlev\'e transcendent $y\approx 6/(\tau_c-\tau)^2$ shows a stronger finite time singularity. A numerical analysis of the one-dimensional back flow model including the full chemical kinetics is welcomed to analyze carefully the finite time singularity.

The saddle-node bifurcation induced DDT is not limited to the ZFK flame model used here in the asymptotic analysis of the one-dimensional back flow model summarized in this manuscript. As already mentioned, the key point is the nonlinear temperature dependence of the laminar flame velocity characterizing any combustible mixture. The latter function can be easily computed numerically for any complex chemical network of real combustible mixtures and used instead of the exponential term in (\ref{yxi}). Moreover the asymptotic analysis can be extended without much difficulty to reduced chemical networks treating the thin zone of radical production as a discontinuity. The results will be qualitatively the same as that presented here.

Such a  DDT mechanism concerns also  turbulent wrinkled flame treated as a chaotic array of elongated flames since a convection flow is systematically induced by the front wrinkling so that the tip of each cell of the cellular of a turbulent flame looks like to that of the elongated flame in {fig.1}.
In the cellular structure of a Rayleigh-Taylor unstable flame expanding freely in spherical geometry, the longitudinal and transverse dimension of the cells are similar $S\approx 1$. In very energetic mixtures the parameter $b\propto  \beta \varepsilon$ (the product of the thermal sensitivity of the flame  velocity by the flame Mach number) might not be very small so that  $S^*\propto 1/(\text e\, b)$ could be close to unity. This could be the case for the nuclear reactions involved in stars.  
Further studies
 are worth performing to decide whether the dynamical saddle-node bifurcation (\ref{sdbif}) is a DDT mechanism explaining the explosion of SNIa supernovae.


\begin{thebibliography}{99}
\expandafter\ifx\csname natexlab\endcsname\relax\def\natexlab#1{#1}\fi

   
\bibitem{shcheltrosh}
K.I. Shchelkin and Yu.K. Troshin, {\em Gasdynamics of Combustion} (Mono Book. corp. Baltimore, 1965).

\bibitem{lee2008}
 J. Lee, {\em The detonation phenomenon} (Cambridge University Press, Cambridge, 2008).

\bibitem{Clavin2016}
P. Clavin and G. Searby, {\em Combustion Waves and Fronts in Flows} (Cambridge University Press, Cambridge, 2016)

  \bibitem{Clavin2023}
 P. Clavin,   "One-dimensional mechanism of deflagration-to-detonation in gas",  J. Fluid. Mech.   Submitted (2023)
 
\bibitem{SaFaw}
A. Sanchez and F.A. Williams, "Recent advances in understanding of flammability characteristics of hydrogen", Progress in Energy and Combustion Science {\bf 41},  1-55 (2014).

\bibitem{ZFK}
Ya.B. Zeldovich and Frank-Kamenetskii, "A theory of thermal flame propagation",   Acta Phys. Chim. {\bf 9},  341-350 (1938).

\bibitem{Wu10}
M-H. Wu, and C-Y. Wang, "Reaction propagation modes in millimeter-scale tubes for ethylene/oxygen mixtures", Proc. Combust. Inst. {\bf 33},  2287-2293
(2011). 

\bibitem{Bikov22}
V. Bikov, A. Koksharov,  M. Kuzetsov, and V.P. Zhukov, "Hydrogen-oxygen flame acceleration in narrow open ended channels", Combust. Flame {\bf 238},  111913
(2022).

\bibitem{Opp66}
P.A. Urtiew and A.K. Oppenheim,  "Experimental observations of the transition to detonation in an explosive gas", Proc. R. Soc. London A {\bf 295},  13-28
(1966).

\bibitem{Lieb10}
M.A. Liberman, M.F. Ivanov, A.D. Kiverin,  M.S. Kuzetsov, A.A Chukalovsky and T.V. Rakhimova, "Deflagration-to-detonation transition in highly reactive combustion mixtures", Acta Astronautica {\bf 67},  688-701
(2010). 

\bibitem{Kuz10}
M. Kuzetsov, M. Liberman and I. Matsukov,  "Experimental study of the preheated zone formation and deflagration to detonation transition", Combust. Sci and Tech. {\bf 182},  1628-1644
(2010).

\bibitem{Zeldo1980}
 Ya. B. Zeldovich, "Regime classification of an exothermic reaction with nonuniform initial condition", Combust. Flame  {\bf 39},  211-214 (1980)
  

\bibitem{Searby96}
 C. Clanet and G. Searby,  "On the tulip flame phenomenon", Combust. Flame  {\bf 105},  225-238 (1996)

\bibitem{Clavin2022}
 P. Clavin,  "Finite-time singularity associated with the deflagration-to-detonation transition on the tip of an elongated flame-front in a tube", Combust. Flame  {\bf 245},  112347 (2022)
 
 \bibitem{Joulin89}
 B. Deshaies and G. Joulin,  "Flame-speed sensitivity to temperature changes and the deflagration-to-detonation transition", Combust. Flame  {\bf 77},  202-212 (1989)
 
 \bibitem{Clavin2021}
 P. Clavin and H. Tofaili,  "A one-dimensional model for deflagration-to-detonation transition on the tip of elongated flames in tubes", Combust. Flame  {\bf 232},  111521 (2021)
 
 \bibitem{Pomeau2012}
 R.D. Peters, M. Le Berre and Y. Pomeau,  "Prediction of catastrophes: a experimental model", Phys. Rev. E  {\bf 86},  026207 (2012)
 

\end{thebibliography}
\end{document}